\documentclass[letterpaper, 10 pt, conference]{ieeeconf}  

\IEEEoverridecommandlockouts                              

\usepackage{graphics} 
\usepackage{epsfig} 
\usepackage{amsmath} 
\usepackage{amssymb}  
\usepackage{xcolor}

\usepackage{booktabs}   
\usepackage{makecell}   
\usepackage{graphicx}   
\usepackage{caption}    

\usepackage{float}

\usepackage{amsmath}   
\usepackage{array}     
\usepackage{booktabs}  
\usepackage{multirow}  
\usepackage{makecell}  
\usepackage{graphicx}  
\usepackage{adjustbox} 

\newcolumntype{P}[1]{>{\centering\arraybackslash}p{#1}}
\newcolumntype{M}[1]{>{\centering\arraybackslash}m{#1}} 

\title{\LARGE \bf
Flatness-based Neural Network Control of DC-DC Converters 
}

\author{
\authorblockN{Kamakshi Tatkare\authorrefmark{1}, Ufuk Topcu\authorrefmark{2}, and Brian Johnson\authorrefmark{1}}
\authorblockA{\authorrefmark{1}Chandra Family Department of Electrical and Computer Engineering,\\
}
\authorblockA{\authorrefmark{2}Department of Aerospace Engineering and Engineering Mechanics, \\}
\authorblockA{The University of Texas at Austin. \\
Email: kamakshi@utexas.edu, utopcu@utexas.edu, b.johnson@utexas.edu}
}

\begin{document}

\maketitle
\thispagestyle{empty}
\pagestyle{empty}

\begin{abstract}

In this work, we present a simple method to engineer a small neural network to control power converters. Learning-based methods have emerged as a promising approach for power converter control over wide operating point variations. However, many of these approaches do not exploit the analytical structure of converter dynamics and, therefore, search over unnecessarily broad policy classes. We address this gap with a systematic framework that combines model-based trajectory generation with policy learning. First, we use differential flatness of averaged converter models to generate feasible state and constrained duty cycle trajectories where duty acts as a control effort signal. Then for policy learning, we use these model-generated trajectories to construct labeled state-to-duty data and train a multilayer perceptron to approximate the control law with a static map. We apply this framework across buck, boost, and buck-boost converters through topology-dependent flat output selection and a common learning pipeline. Simulations show close agreement between learned and model-generated duty trajectories and stable regulation under large synchronous input and load disturbances.
\end{abstract}


\section{INTRODUCTION}
Power electronics circuits, or \emph{converters}, use periodically switched semiconductor devices to shape energy delivery between sources and loads.
The controllable switches and passive components define the converter's dynamical model.
In most power supplies, the control goal is to maintain the load voltage despite disturbances.
The converter is actuated through its switching action, with the \emph{duty ratio}, defined as the fraction of each period that the switch is on, serving as the control effort.
Together, these attributes define a control problem in which voltage regulation is achieved through duty-ratio adjustments~\cite{erickson1982large, banerjee1999nonlinear, sira2006control, kapat2020tutorial}. 

Contemporary learning-based control for power converters spans reinforcement
learning, supervised policy approximation, physics-informed learning,
and adaptive or model-free neural control. Reinforcement learning can
handle nonlinear operation without an explicit plant model, but remains
sensitive to reward design and training configuration; recent
transfer-learning approaches further require multi-stage actor--critic
training and extensive interaction with representative simulation
environments \cite{2025deepRL,chen2024review,shi2026transfer,xu2026qlearning}.
Supervised approaches offer simple online inference, but may rely on
computationally intensive optimization to generate training labels
\cite{nikiforos2025neural}. Model-free neural controllers avoid explicit
converter models by learning from operating data, but shift complexity
to online gradient estimation and parameter adaptation
\cite{model_free}, while neural-network-assisted sliding-mode methods
improve disturbance rejection at the cost of composite observers,
adaptive laws, and fractional-order dynamics \cite{fei2025nnfosmc}.
Physics-informed approaches introduce useful structure, although
energy-based formulations primarily shape energy dynamics rather than
the output-voltage response directly \cite{tatkare2025energy}. More broadly, many learning-based control methods ask the learner to search over a controller class that is broader than the task requires. System structure can therefore be used to narrow the controller search \cite{geva1994tasks}.
 
\textit{Differential flatness} offers a concrete route to trajectory-based control design \cite{fliess1992lessystemesnon,fliess1995flatness}. A flat output provides a parametrization through which the converter states and control input can be written in terms of the output and a finite set of its derivatives. This representation gives an explicit inverse model, allowing a prescribed trajectory to be mapped directly to the corresponding feedforward duty command. Flatness for power electronics appeared first for dc-dc converters in \cite{sira1991differential} and later for dc-ac converters in \cite{sira1999dc}.  
Indirect regulation strategies, which are used for non-minimum phase systems, allowed for flatness to be used on boost and buck-boost circuits~\cite{Fliess1998}. Since then, flatness has been applied to many converter applications. In practice, flatness is commonly paired with feedback strategies such as linearized PI regulation, sliding-mode control, passivity-based designs, and more recently, model predictive control. PI controllers that rely on linearization tend to have reduced disturbance rejection~\cite{sira2006control}.  Sliding-mode controllers suffer from chattering phenomena which in turn require mitigation~\cite{SMC2010}. Passivity-based control, although robust, obscures direct state response shaping~\cite{sira2006control}. Model-based methods like model predictive control are computationally complex \cite{MPC}.  

 In this work, we leverage the structural property of the converter dynamics via flatness. Flatness gives us a direct state-to-duty relation, so we can generate labeled data from the model and train a multilayered perceptron to map that relation. Our contributions are:
\begin{enumerate}
\item A procedure that generates feasible, constrained, and smooth trajectories that encapsulate the desired dynamical response of the system.  
\item A supervised learning based controller that approximates the flatness-based state-to-duty map.
\item Numerical validation of dc-dc converters subjected to large input and load disturbances.
\end{enumerate}

\section{Preliminaries}

\subsection{System Modeling}

\begin{figure*}[t]
  \centering
  \includegraphics[width=0.85\textwidth]{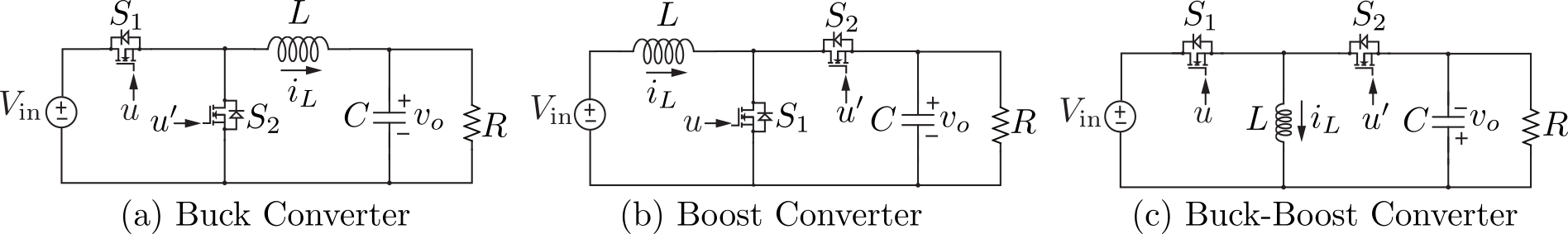}
  \caption{Power stages of ideal basic dc-dc converters.}
  \label{fig:power_stage}
\end{figure*}

The converter dynamics in Fig.~\ref{fig:power_stage} are modeled using the state-space averaging approach of Middlebrook and \'{C}uk~\cite{middlebrook1976general}. Let $\mathbf{x}\in\mathbb{R}^n$ denote the state vector comprising inductor currents, $i_L$, and capacitor voltages, $v_C$, and let $V_{\mathrm{in}}$ denote the input source voltage. The passive elements are represented by the inductance $L$ and capacitance $C$. The control action is the duty ratio $u\in[0,1]$ and its complement $1-u$. The standard form is
\begin{equation}
\begin{aligned}
\dot{\mathbf{x}} &= \big(uA_{1} + (1-u)A_{2}\big)\mathbf{x}
+ \big(uB_{1} + (1-u)B_{2}\big)V_{\mathrm{in}},\\
y_m &= \big(uC_{1}^{\top} + (1-u)C_{2}^{\top}\big)\mathbf{x},
\end{aligned}
\label{eq:ss_avg_final}
\end{equation}
where $y_m$ denotes the measured variable and the matrices $A_{i},B_{i},C_{i}$ are determined by the converter topology.

Equilibrium occurs when the state derivative is zero. Setting $\dot{\mathbf{x}}=\mathbf{0}$ in \eqref{eq:ss_avg_final} yields the steady‑state relation
\begin{equation}
\mathbf{0}
=
\big(u^{\star}A_{1} + (1-u^{\star})A_{2}\big)\mathbf{x}^{\star}
+
\big(u^{\star}B_{1} + (1-u^{\star})B_{2}\big)V_{\mathrm{in}},
\label{eq:steady_state}
\end{equation}
where $u^{\star}$ and $\mathbf{x}^{\star}$ denote the steady‑state duty ratio and state vector, respectively.
The components of $\mathbf{x}^{\star}$ are coupled through the algebraic equilibrium condition \eqref{eq:steady_state}. 

For initial conditions, we assume that the system starts at an initial steady state, i.e., $ \dot{\mathbf{x}}_0=\mathbf{0}.$
Under this assumption, an analogous algebraic relation to \eqref{eq:steady_state} holds at $t=0$
\begin{equation}
\mathbf{0}
=
\big(u_0A_{1} + (1-u_0)A_{2}\big)\mathbf{x}_0
+
\big(u_0B_{1} + (1-u_0)B_{2}\big)V_{\mathrm{in}},
\label{eq:initial_eq}
\end{equation}
which couples the initial state $\mathbf{x}_0$ and the initial duty ratio $u_0$. Given a specified non‑zero initial value for one state component, \eqref{eq:initial_eq} is solved simultaneously with the chosen initial component to determine the remaining initial states and associated control action.

\subsection{Flatness-based Trajectory Generation}

Differential flatness, originally studied by Fliess et al. in
\cite{FLIESS1993159}, is a structural property of certain nonlinear systems. Consider a nonlinear dynamical system with state vector $x\in\mathbb{R}^{n}$ and input control vector $u\in\mathbb{R}^{m}$. Such a system is \emph{differentially flat} if one can define an output vector
$y\in\mathbb{R}^{m}$, known as the \emph{flat output}, whose trajectory contains enough information to reconstruct both the state and input through algebraic relations involving only finitely many time derivatives. More precisely, for some finite $p$,
\[
y = h(x,u,\dot{u},\ldots,u^{(p)})
\]
and there exist functions, for some finite $q$, so that
\[
x = x(y,\dot{y},\ldots,y^{(q)}),
\qquad
u = u(y,\dot{y},\ldots,y^{(q)}).
\]

To identify an appropriate flat output, we examine the internal (zero)
dynamics associated with candidate outputs ~\cite{byrnes1984frequency}. An output is suitable as a
flat variable if, when constrained, the resulting internal dynamics are
well defined and stable. This corresponds to the minimum‑phase property
with respect to that output.

\section{Classical Methods}

Power converters are inherently nonlinear. Engineers usually rely on small-signal methods that linearize the converter model about an operating point. One such simple method, linear state feedback controller has the form, 
\begin{equation}
u
=
-\frac{k_1}{V_{\mathrm{in}}}\sqrt{\frac{L}{C}}\,(i_L-I^\star)
-\frac{k_2}{V_{\mathrm{in}}}\,(v_o-V^\star)
+u^\star 
\end{equation}
for a Boost converter \cite{sira2006control}. The gains, $k1$ and $k2$, are chosen to meet design requirements for a specific operating. 
The converter behaves well compared to the open loop response at this operating point until load resistance and input voltage are change. An integral action will only reduce the steady-state error, and a derivative action can increase damping and reduce oscillations within its limits \cite{han2009pid}. So large-signal methods are needed for robustness \cite{erickson1982large}. 

Classic examples of large signal methods used in power electronics are sliding-mode control \cite{sira2006control} and passivity-based control \cite{rodriguez2001new}. For a boost converter, a sliding surface is reachable using 
\begin{equation}
u = \frac{1}{2}\left(1 - \operatorname{sign}\!\bigl(i_L - i_L^{\mathrm{ref}}\bigr)\right).
\end{equation}
Although the control is robust to both disturbances for $i_L$, it is not robust to regulate $v_o$ to a set-point. Also, this policy leads to chattering phenomena and needs  mitigation techniques \cite{SMC2010}.  Another simple method is the energy-based method presented in \cite{rodriguez2001new} given by \eqref{eq:idapbc}. This static output feedback law is robust against uncertainty in load values but needs adaptive control methods like immersion and invariance to make it robust against changes in input voltage with $v_o$ measured. The focus is on stability and not direct response shaping. The  state feedback exact linearization law proposed in \cite{zheng2012nonlinear}, although robust to both the disturbances when $R$ and $V_\text{in}$ are measured, shapes the transient of 
an energy-like nonlinear function of the current and voltage tracking errors and not output voltage directly. 
\begin{equation}
u=1-\frac{V_{\mathrm{in}}}{V^\star}\left(\frac{v_o}{V^\star}\right)^{\alpha},
\qquad 0 \le \alpha < 1 . \label{eq:idapbc}
\end{equation}
The neural-network-based controllers in~\cite{chan_1993} and \cite{leyva} either employ multiple networks for system identification and adaptive control or require pseudolinearization. Linear methods remain primitive, classical nonlinear methods yield elegant policies but have limits, and contemporary learning-based methods often add needless complexity. We bridge this gap with a simple flatness-informed approach that engineers a small neural network controller.

\section{Proposed Design Method}

\begin{figure}[tbhp]
  \centering
  \includegraphics[width=0.45\linewidth]{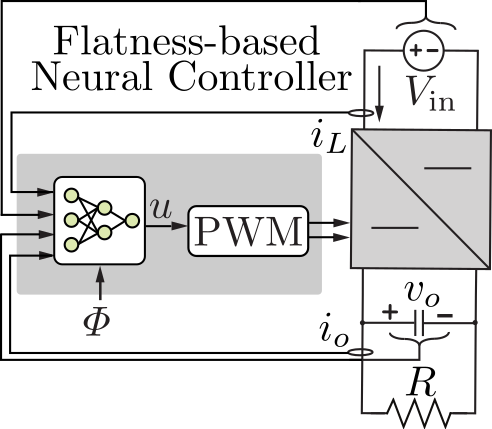}
  \caption{ Proposed  supervised learning trained flatness-based neural controller for dc-dc converters. The inputs to this controller are $i_L$, $v_o$, $R$, and $V_\text{in}$. The pulse width modulator (PWM) generates the duty from the policy $u$.}
  \label{fig:classic}
\end{figure}

We implement the flatness-based control law with a learned feedback map for dc-dc converters as shown in Fig.~\ref{fig:classic}. The disturbance and response shaping procedures follow our prior
work~\cite{tatkare2026inversion}. For a given topology, let the flat output be denoted by \(y(t)\), where the specific choice differs across converters. We have \(y=v_o\) for buck, \(y=i_L\) for boost, and buck-boost as defined in Section~IV. Let
$p(t)\in\{V_{\mathrm{in}}(t),\,R(t)\}$ collect the operating conditions.

\subsection{Disturbance Modeling}
Operating-point changes in converters are naturally modeled as time-varying parameters, e.g., $p(t)\in\{V_{\mathrm{in}}(t),\,R(t)\}$. A sharp change can be written as
\begin{equation}
p(t)=p^{-}+\Delta p\,H(t-t_s),
\end{equation}
where $p^{-}$ and $p^{+}=p^{-}+\Delta p$ are the pre-step and post-step values, $t_s$ is the switching time, and $H(\cdot)$ is the unit step. Since $\frac{d}{dt}H(t-t_s)=\delta(t-t_s)$ we get
\begin{equation}
\dot p(t)=\Delta p\,\delta(t-t_s),
\end{equation}
which induces spikes in reconstructed derivatives and in the computed duty when the flat parameterization with sharp disturbance is differentiated. To avoid impulsive terms, we replace $H(\cdot)$ with a smooth function, 
\begin{equation}
s(t)=\tfrac{1}{2}\left(1+\tanh\!\big(k(t-t_s)\big)\right),
\end{equation}
and model the parameter transition as
\begin{equation}
p(t)=(1-s(t))\,p^{-}+s(t)\,p^{+}, \label{eq:mollifier}
\end{equation}
which yields the bounded derivative 
\begin{equation}
\dot p(t)=\frac{k}{2}\,\mathrm{sech}^2\!\big(k(t-t_s)\big)\,\Delta p.
\label{eq:dmollifier}
\end{equation}
Here $k>0$ controls the transition sharpness. Larger $k$ approaches an ideal step, while smaller $k$ spreads the change over a longer interval.  In real-time control implementations, changes in $R$ and $V_\mathrm{in}$ are typically not ideal step functions, and may be measured or estimated.

\subsection{Response Shaping}

The converters considered in this paper are averaged second-order models, so we shape the flat-output trajectory using a second-order reference template. We select a second‑order, critically damped, reference
transfer function of the form
\begin{equation}
    G(s)=\frac{Y^\star\omega_0^2}{(s+\omega_0)^2},
    \label{eq:tf_G}
\end{equation}
where $\omega_0>0$ is the desired natural frequency. While we choose this for a smooth
settling and minimal overshoot, other standard performance metrics defined on basis of transfer function like settling time, peak time, rise time, and others follow the same procedure.  

Applying a unit‑step reference $1/s$ at the input of \eqref{eq:tf_G}
produces the Laplace transform of the flat output variable,
\begin{equation}
    Y(s) = G(s)\,\frac{1}{s} 
    = \frac{Y^\star\omega_0^2}{s(s+\omega_0)^2}.
    \label{eq:tf_I}
\end{equation}

Taking the inverse Laplace transform yields the time‑domain response as
\begin{align}
    y(t) &= Y^{\star}\Big(1 - e^{-\omega_0 t}\left(1+\omega_0 t\right)\Big).\label{eq:step_response1}
\end{align}
By using the canonical second‑order form \eqref{eq:tf_G} as a design
template, we leverage classical linear system intuition to ensure
desirable dynamics while retaining
the explicit, parameterized structure afforded by differential flatness.

\subsection{Learning-based Control}

The flatness parametrization provides a finite derivative representation of the state variables and the duty ratio in terms of \(y\) and \(p\). In particular, there exists an integer \(r\) and smooth maps \(\Psi_x\) and \(\Psi\) such that
\[
x(t)=\Psi_x\!\left(y(t),\dot y(t),\ldots,y^{(r)}(t);\ p(t)\right),
\]
\[
u(t)=\Psi\!\left(y(t),\dot y(t),\ldots,y^{(r)}(t);\ p(t)\right),
\]
where \(x(t)\) denotes the converter state vector. For second order averaged models \(x=[i_L\ v_o]^{\top}\) and \(u(t)\in[0,1]\).

To generate supervision, we sample multiple operating scenarios by varying \(p(t)\) over the desired ranges and prescribe a smooth reference \(y(t)\). We discretize the flatness parametrization of the system at time steps \(t_k\)  to yield labeled pairs \((z,u)\). With the feedback configuration used in this paper, we define the regressor vector as
\[
z(t)=\begin{bmatrix} x(t)^{\top} & p(t)^{\top} \end{bmatrix}^{\top}\in\mathbb{R}^{n_z}.
\]
The finite dataset, $\mathcal{D}$, of system trajectories, states, disturbances, and control input, takes the form
\[
\mathcal{D}=\left\{\left(z^{(j)},u^{(j)}\right)\right\}_{j=1}^{N},
\qquad u^{(j)}\in[0,1].
\]

We approximate the duty ratio with a neural network \(\pi_{\Phi}:\mathbb{R}^{n_z}\rightarrow[0,1]\),
\[
\hat u=\pi_{\Phi}(z).
\]
Here \(\Phi\) denotes the set of trainable network parameters (weights and biases) of the multilayer perceptron (MLP) \(\pi_\Phi(\cdot)\). We fit \(\Phi\) by  using a mean squared error loss,
\[
\Phi^{\star}=\arg\min_{\Phi}\ \frac{1}{N}\sum_{j=1}^{N}\left(\pi_{\Phi}\!\left(z^{(j)}\right)-u^{(j)}\right)^2.
\]
After training, the online control law evaluates a single forward pass. For enforcing feasibility through saturation, one can implement,
\[
u(t)=\mathrm{sat}_{[0,1]}\!\left(\pi_{\Phi^{\star}}\!\left(z(t)\right)\right).
\]

\section{Application to DC-DC Converters} \label{sec:System Description}
The ideal circuits for the buck, boost, and buck-boost power stages considered in this section are shown in Fig.~\ref{fig:power_stage}, with parasitics ignored. We define the states as the converter output voltage, $v_o$, and the inductor current $i_L$. Let the source voltage be denoted as $V_{\mathrm{in}}$, the load resistance as $R$, and the control input as the duty ratio $u$. The averaged state-space model, equilibrium point, consistent initial conditions, and the selected flat output (based on the stability of the internal dynamics) are listed in Table~\ref{tab:avg_eq_ic_flat}. We use these relations to derive the flat parametrization in this section. A closely related approach appears in \cite{sira1991differential}, where the derivations are based on a normalized averaged model with constant $V_{\mathrm{in}}$ and $R$. Note that the choice of flat variable as Hamiltonian for boost and buck-boost in the same text is not suitable for our approach because of the coupled nature of the states. We use the disturbance models and response shaping described in the previous section to construct the trajectory-generation equations.

\begin{table*}[t]
\centering
\renewcommand{\arraystretch}{1.25}
\setlength{\tabcolsep}{5pt} 
\caption{System Description}
\label{tab:avg_eq_ic_flat}

\small
\begin{adjustbox}{width=\textwidth,center}
\begin{tabular}{M{0.9cm} | M{4.0cm} | *{3}{M{1.25cm}} | *{3}{M{1.25cm}}}
\toprule
\multirow{2}{*}{} &
\multirow{2}{*}{Averaged State Space} &
\multicolumn{3}{c|}{Equilibrium Point} &
\multicolumn{3}{c}{Initial Condition} \\
\cmidrule(lr){3-5}\cmidrule(lr){6-8}
& & $i_L^\star$ & $v_o^\star$ & $u^\star$ & $i_L(0)$ & $v_o(0)$ & $u(0)$ \\
\midrule

\rotatebox[origin=c]{90}{Buck} &
$\begin{aligned}
\dot{i}_L &= -\frac{1}{L}v_o + \frac{u}{L}V_{\mathrm{in}},\\
\dot{v}_o &= -\frac{1}{RC}v_o + \frac{1}{C}i_L
\end{aligned}$ &
$\frac{1}{R}V^\star$ & $V^\star$ & $\frac{V^\star}{V_{\mathrm{in}}}$ &
$\frac{1}{R}V(0)$ & $V(0)$ & $\frac{V(0)}{V_{\mathrm{in}}}$ \\
\midrule

\rotatebox[origin=c]{90}{Boost} &
$\begin{aligned}
\dot{i}_L &= -\frac{1-u}{L}v_o + \frac{1}{L}V_{\mathrm{in}},\\
\dot{v}_o &= \frac{1-u}{C}i_L - \frac{1}{RC}v_o
\end{aligned}$ &
$\frac{(V^\star)^2}{RV_{\mathrm{in}}}$ & $V^\star$ & $1-\frac{V_{\mathrm{in}}}{V^\star}$ &
$\frac{(V(0))^2}{RV_{\mathrm{in}}}$ & $V(0)$ & $1-\frac{V_{\mathrm{in}}}{V(0)}$ \\
\midrule

\rotatebox[origin=c]{90}{\makecell{Buck-\\Boost}} &
$\begin{aligned}
\dot{i}_L &= -\frac{1-u}{L}v_o + \frac{u}{L}V_{\mathrm{in}},\\
\dot{v}_o &= -\frac{1}{RC}v_o + \frac{1-u}{C}i_L
\end{aligned}$ &
$\frac{V_{\mathrm{in}}+V^\star}{V_{\mathrm{in}}R}V^\star$ & $V^\star$ & $\frac{V^\star}{V^\star+V_{\mathrm{in}}}$ &
$\frac{V_{\mathrm{in}}+V(0)}{V_{\mathrm{in}}R}V(0)$ & $V(0)$ & $\frac{V(0)}{V(0)+V_{\mathrm{in}}}$ \\
\bottomrule
\end{tabular}
\end{adjustbox}
\end{table*}

\subsection{Buck Converter} 

 A buck converter is a dc-dc converter whose control objective is to regulate $v_o < V_\mathrm{in}$.
With $y=v_o$ chosen as the flat output, we express the
remaining state and input trajectories directly in terms of $y$ and its derivatives. Solving for $i_L$ using the equations in  Table~\ref{tab:avg_eq_ic_flat} gives
\begin{align}
i_L &= C\,\dot{y} + \frac{1}{{R}}\,y
\label{eq:buck_flat_param_iL}
\end{align}
Substituting the derivative of \eqref{eq:buck_flat_param_iL} into the state space equation and solving for the duty ratio
yields
\begin{align}
u
&= \frac{L}{{V_\mathrm{in}}}\,{\left(C\,\ddot{y} + \frac{1}{R}\,\dot{y}\right)} + \frac{1}{{V_\mathrm{in}}}\,y
\label{eq:flat_param_u}
\end{align}

These equations parameterize the states and control but have $ V_\text{in}$ and $R$ constants. To learn the dynamic response of the converter model when physical parameters change abruptly we now consider the effect of a sharp step change in the load resistance, from ${R}^{-}$ to ${R}^{+}$. Differentiating the current expression, \eqref{eq:buck_flat_param_iL}, yields
\begin{align}
\dot{i}(t) &= C\,\ddot{y}(t) + \frac{\dot{y}(t)}{R(t)} - \frac{y(t)\,\dot{R}(t)}{R(t)^2}, \label{eq: buck_flat_i_mollified}
\end{align}
Under this change, the inductor current exhibits an instantaneous jump
revealing an impulsive term in the current derivative. Substituting this into the duty ratio expression shows that the
duty ratio inherits a non-desirable impulse. A sharp step in the input
voltage, which, while not introducing an impulse in the current dynamics, produces a finite jump in $u(t)$ at $t_s$, again leading to numerical stiffness.


Hence, we use \eqref{eq:mollifier} and \eqref{eq:dmollifier} to model $V_{\mathrm{in}}(t)$  and $R(t)$.  We substitute \eqref{eq: buck_flat_i_mollified} into equation the state space equation and solve the duty ratio to obtain
\begin{align}
u
&= \frac{L}{V_\text{in}(t)}
\left(
C\,\ddot{y}(t)
+ \frac{\dot{y}(t)}{R(t)}
- \frac{y(t)\,\dot{R}(t)}{R(t)^2}
\right)
+ \frac{y(t)}{V_\text{in}(t)} .
\label{eq:flat_mollifier_param_u}
\end{align}


\subsection{Boost}
 A boost converter is a dc-dc converter whose control objective is to regulate $v_o > V_\mathrm{in}$.
We choose $y=i_L$ as the flat output as it leads to stable internal dynamics.  We express the remaining state and input trajectories directly in terms of $y$ and its derivatives. Solving for $i_L$ using the equations in  Table~\ref{tab:avg_eq_ic_flat} gives
\begin{align}
\dot{u} &=
\frac{
(1-u)RLC\,\ddot{y}
- {(1-u)\left({V_\mathrm{in}} - L\,\dot{y}\right)}
+ (1-u)^3 {Ry}
}{
RC({V_\mathrm{in}} - L\,\dot{y})
}, \label{eq:boost_udot}\\
v_o &= \frac{{V_\mathrm{in}} - L\,\dot{y}}{1-u} \qquad \text{for } u \neq 1. \label{eq:boost_v_from_y}
\end{align}
From \eqref{eq:boost_udot} and \eqref{eq:boost_v_from_y} it is evident that the parameterization is not pure. We use prolonged actuation by letting the control action evolve with time and become a new state for the system. This method is described in \cite{levine2023differential}. 

To learn the response of the converter model when physical parameters alter abruptly we include $R(t)$ and $V_\text{in}(t)$. Since the steady state value of current depend on input-source voltage and load resistance, an interesting series of equations results for the transition dynamics,  with non-zero denominators,
\begin{align}
I^{\star}(t) &= \frac{\left(V^{\star}\right)^2}{R(t)\,{V_\mathrm{in}}(t)}, \qquad I^{\star}(0) = \frac{\left(V^{\star}\right)^2}{R(0)\,{V_\mathrm{in}}(0)}, \nonumber\\
U^{\star}(t) &= 1-\frac{{V_\mathrm{in}}(t)}{V^{\star}}, \qquad
U^{\star}(0)=1-\frac{V_\mathrm{in}(0)}{V(0)}. \label{eq:Ustar_t}
\end{align}
We introduce a new variable $\phi(t)$, a unit gain critically damped response, for mathematical ease. 
\begin{align}
\phi(t) &\triangleq 1-e^{-\omega_0 t}\left(1+\omega_0 t\right) \label{eq:phi_ddot} 
\end{align}
The flat variable in terms of $\phi(t)$ and time varying $I^{\star}(t)$ is
\begin{align}
y(t) &= I^{\star}(t)\,\phi(t), \label{eq:y_t}
\\
\dot{y}(t) &= \dot{I}^{\star}(t)\,\phi(t) + I^{\star}(t)\,\dot{\phi}(t), \label{eq:y_dot_t}\\
\ddot{y}(t) &= \ddot{I}^{\star}(t)\,\phi(t) + 2\,\dot{I}^{\star}(t)\,\dot{\phi}(t)
+ I^{\star}(t)\,\ddot{\phi}(t). \label{eq:y_ddot_t}
\end{align}
For \eqref{eq:y_dot_t} and \eqref{eq:y_ddot_t} we find first and second derivative of $I^\star$, \eqref{eq:Ustar_t}, in \eqref{eq:Istar_dot} and \eqref{eq:Istar_ddot} respectively. 
\begin{align}
\dot{I}^{\star}(t) &= -\left(V^{\star}\right)^2\,
\frac{R(t)\,\dot{V_\mathrm{in}}(t) + \dot{R}(t)\,V_\mathrm{in}(t)}{\big(R(t)V_\mathrm{in}(t)\big)^{2}},
\label{eq:Istar_dot}\\[6pt]
\ddot{I}^{\star}(t)
&=
\frac{\left(V^{\star}\right)^2}
{\big(R(t)V_\mathrm{in}(t)\big)^3}
\Big(
R(t)^2
\big(
2\dot{V}_\mathrm{in}(t)^2
-
V_\mathrm{in}(t)\ddot{V}_\mathrm{in}(t)
\big)
\nonumber\\
&\quad
+
V_\mathrm{in}(t)^2
\big(
2\dot{R}(t)^2
-
R(t)\ddot{R}(t)
\big)
\nonumber\\
&\quad
+
2R(t)V_\mathrm{in}(t)\dot{R}(t)\dot{V}_\mathrm{in}(t)
\Big). \label{eq:Istar_ddot}
\end{align}
To model smooth functions as transitions in time‑varying parameters, we follow \eqref{eq:mollifier}. For the load
resistance, we define
\begin{align}
R(t) &= (1-s(t))\,R^{-} + s(t)\,R^{+}, 
\label{eq:R_ddot_mollifier}
\end{align}
where $\Delta R=R^{+}-R^{-}$ and $k>0$ governs the transition sharpness. Similarly, for a general disturbance parameter $V_\mathrm{in}(t)$ with pre‑ and
post‑event values $V_\text{in}^{-}$ and $V_\text{in}^{+}$, we define
\begin{align}
V_\mathrm{in}(t) &= (1-s(t))\,V_\mathrm{in}^{-} + s(t)\,V_\mathrm{in}^{+},  \label{eq:E_ddot_mollifier}
\end{align}
where $\Delta V_\mathrm{in} = V_\mathrm{in}^{+}-V_\mathrm{in}^{-}$.

Finally, we derive prolonged parameterization using the above definitions and average state space model. The equation \eqref{eq:boost_udot} for control action becomes \eqref{eq:udot_tv} and the equation \eqref{eq:boost_v_from_y} for voltage state becomes \eqref{eq:v_rec_tv}. 
\begin{align}
\dot{u} &= \frac{1-u}{V_\mathrm{in} - L\,\dot{y}}
\left(
L\,\ddot{y} + \frac{L}{RC}\,\dot{y} + \frac{(1-u)^2}{C}\,y
- \dot{V_\mathrm{in}} - \frac{V_\mathrm{in}}{RC}
\right), \label{eq:udot_tv}\\[6pt]
v_o &= \frac{V_\mathrm{in} - L\,\dot{y}}{1-u} \qquad \text{for } u \neq 1. \label{eq:v_rec_tv}
\end{align}
\subsection{Buck-Boost}
A buck-boost converter is a dc-dc converter whose control objective is to regulate $v_o < V_\mathrm{in}$ or $v_o > V_\mathrm{in}$ while inverting the polarity.
We choose $y=i_L$ as the flat output and parametrize $v_o$ and $u$ directly in terms of $y$ and its derivatives. We solve the equations in  Table~\ref{tab:avg_eq_ic_flat} to get

\begin{align}
\dot{u} &=
\frac{(1-u)\!\left(L R C\,\ddot y \;+\; L\,\dot y \;+\; R \,(1-u)^{2}y \;-\; u\,V_{\mathrm{in}}\right)}
{R C\,\left(V_{\mathrm{in}} - L\,\dot y\right)}, \label{eq:bb_udot}\\
v_o &= \frac{u\,V_{\mathrm{in}} - L\,\dot y}{1-u} \qquad \text{for } u \neq 1
\label{eq:bb_v_from_y}
\end{align}
From \eqref{eq:bb_udot} and \eqref{eq:bb_v_from_y} it is evident that the parametrization is not pure. Once more we use prolonged actuation for the system as described in \cite{levine2023differential}. 
For abrupt changes in $R$ and $V_{\mathrm{in}}$ we follow the boost case. The steady state value of current depends on non-zero input-source voltage and non-zero load resistance, an interesting series of equations results for the transition dynamics, with $v_{o}^* =  V^*$
\begin{align}
I^{\star}(t) &= \frac{V^{\star}\bigl(V_{\mathrm{in}}(t)+V^{\star}\bigr)}{R(t)\,V_{\mathrm{in}}(t)},
\quad U^{\star}(t) = \frac{V^{\star}}{V_{\mathrm{in}}(t)+V^{\star}}. 
\end{align}
We use $\phi(t)$ defined in \eqref{eq:phi_ddot} to express flat variable in terms to this variable just like we did in \eqref{eq:y_t} to \eqref{eq:y_ddot_t}.
Again for $\dot y$ and $\ddot y$, we find first and second derivative of $I^\star$ which involves finding derivatives of $R$ and $V_\mathrm{in}$.


To model smooth functions as transitions in time‑varying parameters, we follow \eqref{eq:mollifier} and finally, we derive prolonged parameterization with time varying disturbances using the above definitions and average state space form. The equation \eqref{eq:bb_udot} for control action becomes \eqref{eq:bb_udot_tv} and the equation \eqref{eq:bb_v_from_y} for voltage state becomes \eqref{eq:bb_v_rec_tv}. 
\begin{align}
\dot u &= 
\frac{(1-u)}{R(t)\,C\,\bigl(V_{\mathrm{in}}(t) - L\,\dot y\bigr)}
\Bigl(L\,R(t)\,C\,\ddot y + L\,\dot y \nonumber\\[-2pt]
&\quad +R(t)\,(1-u)^{2}\,y - u\,V_{\mathrm{in}}(t)
- R(t)\,C\,u\,\dot V_{\mathrm{in}}(t)\Bigr),
\label{eq:bb_udot_tv}\\
v_o &= \frac{u\,V_{\mathrm{in}}(t) - L\,\dot y}{\,1 - u\,} \qquad \text{for } u \neq 1.
\label{eq:bb_v_rec_tv}
\end{align}

\section{Numerical Results}\label{sec:results}

\begin{figure*}[!t]
  \centering
  \includegraphics[width=0.9\textwidth]{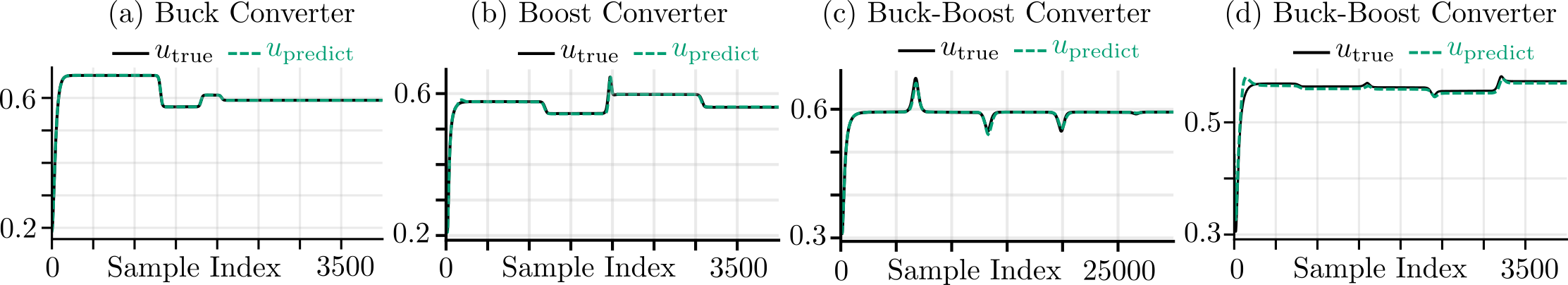}
  \caption{Validation of the learned duty map on unseen trajectories. }
  \label{fig:model_val}
\end{figure*}

\begin{figure*}[!t]
  \centering
  \includegraphics[width=0.95\textwidth]{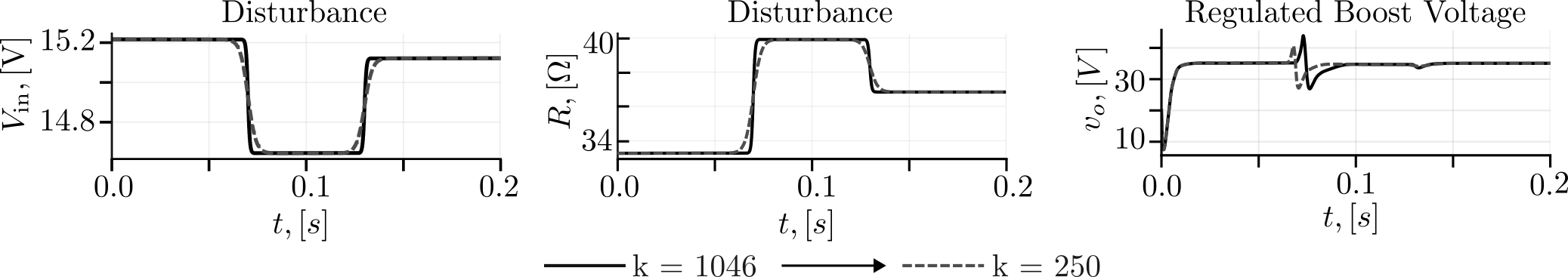} 
  \caption{Effect of the smoothing parameter \(k\) on closed-loop response for the boost converter. 
}
  \label{fig:k_smooth}
\end{figure*} 

\begin{figure*}[!t]
  \centering
  \includegraphics[width=0.95\textwidth] {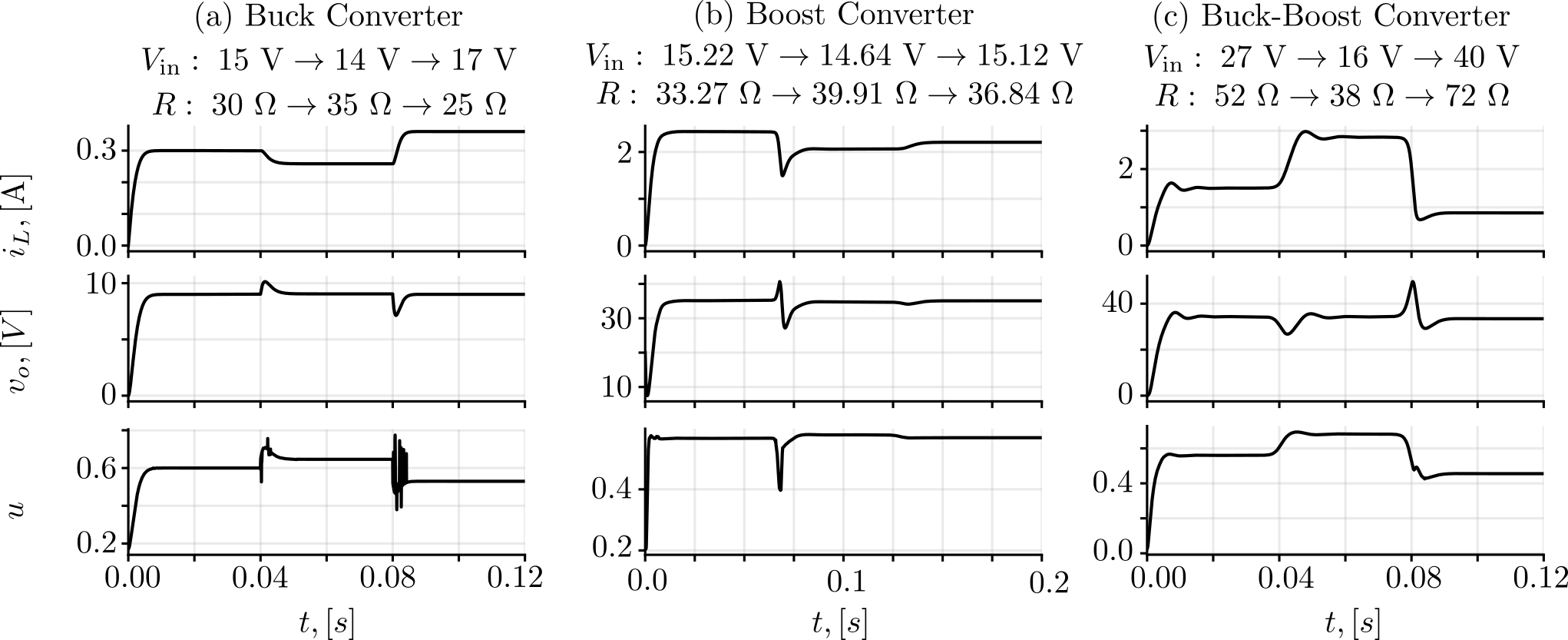}
  \caption{Closed-loop response to smooth variation in output resistance and input voltage at the same time.}
  \label{fig:baseline_responses}
\end{figure*}

\subsection{Data-generation and Training Setup}

We construct the data set by applying the flatness-based parameterization developed in Section~\ref{sec:System Description} across a wide range of operating conditions and transient scenarios. Specifically, we sample state operating points, initial conditions, and disturbance profiles using Latin hypercube sampling \cite{mckay2000comparison}. We construct smooth trajectories, $N_\text{traj}$, using the shaping parameters, $\omega_0$, and $k$, listed in Table~\ref{tab:params}. The generated samples are organized as input-output pairs \((z,u)\), where \(z\) contains the feedback states together with the operating conditions. Finally, we use
\[
z=\begin{bmatrix} i_L & v_o & V_{\mathrm{in}} & R \end{bmatrix}^{\top}, \qquad u\in[0,1].
\]
The dataset is split into training, testing, and validation set. Model is trained with batch-size of size 32.  Table~\ref{tab:converter_summary}, reports converter specific activation function, layer widths, and total number of trainable parameters of the MLP. We use the Adam optimizer with the learning rate and early-stopping epoch given in Table~\ref{tab:converter_summary}. All models are implemented in PyTorch on an Intel CPU (Xeon W9-3495X) workstation with an NVIDIA RTX 6000 Ada GPU.

\begin{table}[ht]
  \centering
  \caption{Nominal parameter values}
  \label{tab:params}
  \begin{tabular}{lccccccc}
    \toprule
    Converter
      & \makecell{$L$\\$[\mathrm{mH}]$}
      & \makecell{$C$\\$[\mu\mathrm{F}]$}
      & \makecell{$V_{\rm in}$\\$[V]$}
      & \makecell{$V^*$\\$[V]$}
      & ${k}$
      & $\omega_o$
      & \makecell{$N_{\mathrm{traj}}$} \\
    \midrule
    Boost
      & 20  & 20  & 15  & 35  & 1046  & 650  & 150 \\
    \addlinespace
    Buck
      & 20  & 20  & 15  & 9   & 11000 & 850  & 50 \\
    \addlinespace
    Buck--Boost
      & 10  & 20  & 24  & 35  & 500   & 500  & 500 \\
    \bottomrule
  \end{tabular}
\end{table}

\begin{table*}[!t]
  \centering
  \caption{Dataset generation and training details for converter topologies}
  \label{tab:converter_summary}
  \begin{tabular}{lccc ccccc c}
    \toprule
    Topology 
      & \makecell{$V_{\rm in}$ range\\$[V]$}
      & \makecell{$R$ range\\$[\Omega]$}
      & \makecell{Total\\samples}
      & \makecell{Activation\\function}
      & \makecell{Neurons/\\layer}
      & \makecell{Trainable\\parameters}
      & \makecell{Test MSE - Epoch}
      & \makecell{Learning \\rate}\\
    \midrule
    Boost
      & $[14.5, 15.5]$   & $[32, 40]$    & 600,000
      & \texttt{sigmoid}     & 256–128–64
      & 42,497       & $2.1\times 10^{-5}, 318$     & $0.001$ \\
    \addlinespace
    Buck
      & $[13, 18]$   & $[25, 35]$    & 200,000
      & \texttt{tanh}     & 64–64–64
      & 8,705       & $0 \times 10^{0}, 48$      & $0.001$\\
    \addlinespace
    Buck–Boost \#1
      & $[24]$   & $[20, 40]$    & $16 \times 10^{6}$
      & \texttt{tanh}     & 64–64–64
      & 8,641       & $2 \times 10^{-5}$, 100      & 0.0001 \\
        \addlinespace
    Buck–Boost \#2
      & $[20,50]$   & $[20,75]$    & $2 \times 10^6$
      & \texttt{sigmoid}     & 256–128–64
      & 42,497       & $2.3 \times 10^{-5}, 5$      & 0.001 \\
    \bottomrule
  \end{tabular}
\end{table*}

\subsection{Model Inference}

Figure~\ref{fig:model_val} validates the learned duty map on trajectories not used for training. The duty ratio from the flatness-based construction, $u_{\mathrm{true}}$, and the network output, $\hat{u}=u_{\mathrm{predict}}$, nearly coincide for the buck, boost, and buck-boost converters (\#1 - boost mode in (c) and \#2 - buck-boost mode in (d)). Small mismatch occurs in (d), where the network learns to perform both step-up and step-down operation as opposed to the boost operation in (c). The validation loss are reported in Table~\ref{tab:converter_summary}.

Figure~\ref{fig:k_smooth} shows how the smoothing parameter $k$ shapes the boost converter response under step changes in $V_{\mathrm{in}}$ and $R$. Smaller $k$ spreads the transition over a longer interval and yields smoother duty action. Larger $k$ produces a steeper transition and increases transient peaking in $i_L$ and $v_o$.

Figure~\ref{fig:baseline_responses} reports closed-loop time responses under simultaneous changes in $V_{\mathrm{in}}$ and $R$. For the buck and buck-boost converters the operating point changes at $t=0.04\,\mathrm{s}$, and $t=0.08\,\mathrm{s}$, while for the boost converter it changes at $t=0.066\,\mathrm{s}$ and $t=0.133\,\mathrm{s}$. In each case, the duty ratio adjusts quickly and the states settle to the new steady values after a short transient. Jitter in the duty corresponds to the inference rate of MLP.

\section{CONCLUSIONS}
This paper presented a flatness-based learning framework for nonlinear control of averaged dc-dc converter models. We generate feasible state and duty trajectories offline by explicit flatness parametrization with smooth transition shaping, then distill the resulting duty law into a static neural map. Numerical studies on buck, boost, and buck–boost converters show that the learned duty closely matches the model-generated duty and maintains regulation under large input and load changes. This study demonstrates feasibility on averaged converter models which exhibit the structural property of pure flatness or flatness by prolongation. Future work will develop formal stability guarantees and implement the controller in real-time. 

\addtolength{\textheight}{-12cm}   



\section*{ACKNOWLEDGMENT}
This research was supported in part by the National Science Foundation award number 2409535, ONR N00014-21-1-2164, and a generous gift from Texas Instruments. The authors acknowledge the use of ChatGPT-5.4 for editorial assistance and code formatting. The authors remain responsible for all material presented in this manuscript.


\bibliographystyle{ieeetr}
\bibliography{bibliography_clean}
\end{document}